\documentclass[10pt,twocolumn]{article}
\usepackage[cp1251]{inputenc}
\usepackage[english]{babel}
\usepackage{graphicx}
\usepackage{amssymb}
\usepackage{amsmath}
\usepackage{amsthm}
\usepackage{amsfonts}
\usepackage{amssymb}
\title{{\bf \Large Exact Cosmological Models with the Yang - Mills Fields \\on Lyra Manifold}\\
{\normalsize ~~{\bf V.\,K. Shchigolev\thanks{E-mail:
vkshch@yahoo.com}\,,\, D.\,N. Bezbatko}}\\
{\small {\it Department of Theoretical Physics, Ulyanovsk State University,
Ulyanovsk, 432000, Russia}}\\
\vspace{2mm}
\small \begin{quote}{\bf Abstract} --  The present study deals with the Friedmann-Robertson-Walker cosmological models of the Yang-Mills (YM) fields in Lyra geometry. The energy-momentum tensor  of the YM fields for our models is obtained with the help of exact solution for the corresponding YM equations subject to their minimal coupling to gravity. Two specific exact solutions of the model are obtained regarding the effective equation of state and the exponential law of expansion. Some physical and geometrical behavior of the model are also discussed.
 \\
\vspace{2,5mm}
{\bf PACS numbers}: 98.80.-k ; 98.80.Jk; 04.50.Kd; 04.20.Jb; 04.40.-b
 \\
{\bf Key words}: Cosmological model; Lyra geometry; Yang-Mills fields.\\
\end{quote}}
\date{}

\begin{document}

\maketitle
\vspace{-2.5cm}
\section{Introduction}

The relatively  recent discovery of the present acceleration of our Universe is strictly proved by Type Ia Supernovae, Cosmic Microwave Background Radiation, and Sloan
Digital Sky Survey \cite{Riess}-\cite{Allen}.  This fact initiated a great number of theoretical hypothesis, and inspired researchers to a diversity of different explanations  of such an unusual behavior of the Universe. Nevertheless, mostly two distinct approaches to explain the phenomenon of accelerated expansion of the Universe have been well developed during the recent years. One of them deals with the so-called dark energy (DE) with negative pressure in General Relativity, another one acts within the frameworks of some acceptable modifications of the gravity theory on the long distance.

According to several estimations, the dark energy occupies up to 72 \% of the total energy of the universe,
while dark matter gives about 24 \%,  and the usual baryonic matter consists of about 4 \% of the total energy.
In spite of all attempts to explain the late acceleration of our Universe, DE is still an open problem to the theoretical physicists because its nature is unknown so far. As known, there are numerous models available in the literature devoted to different descriptions of the nature of DE but none of them are fully definitive and reliable.  The simplest and the most natural candidate
among them is the cosmological constant \cite{Peebles} with the equation of state (EoS) parameter
$w_{\Lambda} =-1$. Although this model meets the observational data but it suffers at least from two serious problems such as fine tuning and cosmic coincidence.

Currently, there are numerous DE and modified gravity models to explain the present acceleration of the universe.  Alternatively, to explain
the decay of the density, the different forms of dynamically changing DE with an effective
equation of state (EoS) $w_{de} = p_{de}/\rho_{de} < -1/3$, were proposed instead of a constant
vacuum energy density. Other possible forms of DE include quintessence $-1/3> w_q > -1$, phantom $(w_{ph} < -1)$  etc.

In the last few decades there has been considerable interest in alternative theories of gravitation. One of the most interesting among them is the scalar-tensor theory proposed by G. Lyra \cite{Lyra}, and essentially developed by Sen and Dunn \cite{Sen}. They have constructed  an analogue of the
Einstein field equation based on Lyra's geometry.
According to Halford \cite{Halford} the
scalar-tensor treatment based on Lyra's geometry
predicts the same effects within observational
limits as in Einstein's theory.  Moreover, as it pointed out by Soleng \cite{Soleng}, the constant displacement field in
Lyra's geometry will either include a creation field
and be equal to Hoyle's creation field cosmology
or contain a special vacuum field, which together
with the gauge vector form may be considered as a
cosmological term. Subsequent investigations were
done by several authors in cosmology  within the framework of Lyra's
geometry (see, e.g., \cite{Shchigolev1}-\cite{Shchigolev6} and references therein).

Involving of the exotic forms of matter and fields for the explanation of the phenomenon of accelerated expansion in the context of Lyra's geometry is reasonable in the same extent as it is in Einstein's theory of gravity. Thereby the models on Lyra manifold with the quintessence, phantom field, tachyons, Chaplygin gas and some other forms of matter were studied by several authors. Nevertheless, the problem which has to be investigated in detail still remains unsolved, namely, whether the resource of Lyra's geometry in the form of the displacement field is quite enough to describe the acceleration of the universe filled by the ordinary matter and fields.

As known, the electromagnetic fields could be one of these natural sources of gravity . A number of studies in the framework of Lyra's geometry involved the electromagnetic fields as the sources of gravity in the cosmological models (see, e.g., \cite{Pradhan}-\cite{Ali} and references therein).
As far as we know, the models in Lyra's geometry  which include the Yang-Mills fields \cite{Yang} have not  been studied yet. At the same time, a large  number of works investigated the influence and role of these fields on the behavior of the Universe in the framework of Einstein's theory (see, e.g. \cite{Shchigolev3}-\cite{Gal} to mention just a few).

Motivated by the above mentioned investigations, we have considered the homogeneous  cosmological model with the Yang-Mills field in Lyra geometry. The aim of this paper is to search for the exact solutions for such a model, that is to find explicit expressions for the scale factor and the displacement field of this model.  As consequence, through the time dependence of the YM content of the Universe, solutions are found
for the energy density and the pressure. Some geometrical and physical aspects of the model also have  been studied. For instance, exact solutions for the scalar field and its potential have been obtained in two specific cases regarding the effective equation of state and the exponential law of expansion.

%==============================
\section{The model equations}
%==============================

The Einstein's gravitational equations in Lyra's geometry  in normal gauge (see, e.g. \cite{Sen} and \cite{Matyjasek1}) can be written as
\begin{equation}\label{1}
R_{ik}- \frac{1}{2} g_{ik} R - \Lambda g_{ik} +  \frac{3}{2}\phi_i \phi_k - \frac{3}{4}g_{ik}\phi^j \phi_j = T^{(tot)}_{ik},
\end{equation}
where $\phi_i$ is a displacement vector, and  $\Lambda$ is a time-varying cosmological term. For simplicity, we assume  the gravitational constant $8\pi G=1$. All other symbols have their usual meanings in the Riemannian geometry, and $T^{(tot)}_{ik}$ stands for the total energy-momentum tensor (EMT) of all kinds of matter in the Universe.
The Friedmann-Robertson-Walker (FRW) line element can be represented by
\begin{equation}
ds^2 = d t^2- a^2 (t)[d r^2+\xi^2(r)d \Omega^2], \label{2}
\end{equation}
where $a(t)$ is a scale factor of the Universe, and
\begin{eqnarray}
\xi_(r)=\left\{\begin{array}{rcl}
\sin(r),~~~~&&k =+1,\\
r,~~~~~~~~~~&&k =0,\\
\sinh(r),~~&&k =-1.\\
\end{array}
\right.
 \label {3}
\end{eqnarray}
in accordance with the  sign of the spatial  curvature $k$.

In the case of minimal coupling of the displacement field with matter, the energy-momentum tensor of matter can be derived in a usual manner from the Lagrangian of matter. Therefore, the EMT of the YM fields  is followed from the Lagrangian density of $SO_3$ YM field \cite{Yang} as
\begin{equation}\label{4}
L_{YM}=-\frac{1}{16\pi}F_{ik}^{a}F^{a ik},
\end{equation}
where $F_{ik}^{a}=\partial_{i}W_{k}^{a}-\partial_{k}W_{i}^{a}+
e\varepsilon_{abc}W_{i}^{b}W_{k}^{c}$ is the stress tensor of YM fields $W^a_i$.
Varying this Lagrangian with respect to metric, one can easily get
the following EMT of YM fields
\begin{equation}\label{5}
T_{i}^{k}=-\frac{1}{4\pi}F_{ij}^a
F^{akj}+\frac{1}{16\pi}\delta_{i}^{k} F_{mn}^{a}
F^{amn}.
\end{equation}
At the same time, the variation of Lagrangian (\ref{4}) with respect to YM fields $W^a_i$ yields the following YM equation
\begin{equation}\label{6}
D_{i}\left(\sqrt{-g}F^{aik}\right)=0,
\end{equation}
where $D_{i}$ stands for the covariant derivative.

Let us suppose further that the remaining content of the universe can be described as a perfect fluid with the EMT of the following form
\begin{equation}\label{7}
\mathcal{T}_{ik}= (\rho_m +p_m)u_i u_k -p_m\, g_{ik},
\end{equation}
where  $u_i = (1,0,0,0)$ is  4-velocity of the co-moving observer, satisfying $u_i u^i = 1$, $\rho_m$ and $p_m$ are the energy density and pressure of matter, consequently. Hence, the total EMT equals $T^{(tot)}_{ik}=T_{ik}+\mathcal{T}_{ik}$.

%==============================
\section{Exact solution for the YM equation}
%==============================

As well-known, the generalized Wu-Yang {\it ansatz} for the $SO_3$ YM
fields can be written as \cite{Samaroo}
\begin{eqnarray}\label{8}
W^a_{\mu} =\varepsilon_{\mu ab}x^b\frac{K(r,t)-1}{gr^2}+\Bigl(\delta^a_{\mu}-\frac{x^ax_{\mu}}{r^2}
\Bigr)\frac{S(r,t)}{gr}, \nonumber \\
W^a_0 = x^a \frac{W(r,t)}{gr},~~~~~~~~~~~~~~~~~~~~~~~~~~~~~~~~~~~~~~~~
\end{eqnarray}
where $\mu=1,2,3$. The example of exact solution for the YM equation (\ref{6}) in metric (\ref{2}) has been obtained in \cite{Samaroo} with the help of the following substitution
\begin{equation}\label{9}
W = \frac{d \alpha}{d t}, ~~K = P(r)\cos \alpha(t),~~ S = P(r)\sin \alpha(t),
\end{equation}
where $\alpha$ is an arbitrary function of time.

As a result of simple calculation on the base of (\ref{8}) and (\ref{9}), we have the following expressions for the YM tensor components
\begin{eqnarray}
\left\{\begin{array}{rcl}
{\bf F}_{12}=g^{-1} P'(r)\Bigl({\bf m }\,\cos\alpha + {\bf
l}\,\sin\alpha\Bigr),~~~~~~~~~~~~~~~~~\\
{\bf F}_{13}=g^{-1} P'(r)\sin\theta\Bigl({\bf m }\,\sin\alpha-{\bf
l}\,\cos\alpha\Bigr),~~~~~~~~~~~\\
{\bf F}_{23}=g^{-1}\sin\theta\Bigl(P^2(r)-1\Bigr){\bf n},~~~~~~~~~~~~~~~~~~~~~~~~~ \\
{\bf F}_{01}={\bf F}_{02}={\bf F}_{03}=0, ~~~~~~~~~~~~~~~~~~~~~~~~~~~~~~~~~~~\\
\end{array}
\right.\label{10}
\end{eqnarray}
where
\begin{eqnarray}
{\bf n}= (\sin\theta \cos \phi, \sin \theta \sin \phi,
\cos \theta),~~~~~~~~\nonumber
\\
{\bf l} = (\cos\theta \cos \phi, \cos \theta \sin
\phi, -\sin \theta),~~~~~\nonumber
\\
{\bf m}= (-\sin \phi, \cos \phi, 0)~~~~~~~~~~~~~~~~~~~~~~
\nonumber
\end{eqnarray}
are the orthonormalized isoframe vectors, and the prime means a
derivative with respect to $r$. As it is noted in \cite{Samaroo}, the YM fields (\ref{10}) have only magnetic components.

Here and below an overdot denotes the partial derivative with respect to time $t$,
and a prime denotes the partial derivative with respect to $r$.
Taking into account the Wu-Yang {\it ansatz} (\ref{10}), the energy-momentum tensor (\ref{4}) can be written down in the form
\begin{eqnarray}
T_0^0 = \frac{1}{8\pi g^2 a^4}\left(a^2 W'^2+a^2 Z + X + Y\right),~~~~~~~~~~~~~~~~\nonumber
\\
T_1^1 =\frac{1}{8\pi g^2 a^4}\left(a^2 W'^2-a^2 Z - X + Y\right),~~~~~~~~~~~~~~~~\label{11}
\\
T_2^2 = T_3^3 = - \frac{1}{8\pi g^2 a^4}\left(a^2 W'^2+ Y\right),~T_0^1=\frac{1}{2\pi g^2 a^2}J,~\nonumber
\end{eqnarray}
where
\begin{eqnarray}
X=\frac{2(K'^2+S'^2)}{\xi^2(r)},~Y=\frac{(K^2-1+S^2)^2}{\xi^4(r)},\nonumber\\
Z=\frac{2[(\dot K+WS)^2+(\dot S-WK)^2]}{\xi^2(r)},~~~~~~~~~\label{12}
\\
J=\frac{K'(\dot K+WS)+S'(\dot S-WK)}{\xi^2(r)}.~~~~~~~~~\nonumber
\end{eqnarray}

The set of equations for the YM fields (\ref{6}) can be written down with the help of
{\it ansatz} (\ref{8}) as follows
\begin{eqnarray}
\left(\xi ^2 W'\right)'
+2\left[\left(\dot S-WK \right)K-\left(\dot K+WS \right)S\right]=0,  \nonumber\\
a\frac{\partial}{\partial t}\left[a\left(\dot S-WK\right)\right]-
S'' ~~~~~~~~~~~~~~~~~~~~~~~~~~~~~~~~\nonumber\\
+\left[\frac{\left(K^2-1+S^2 \right)S}{\xi^2}
-\left(\dot K+WS \right)W\right]=0,\label{13}\\
a\frac{\partial}{\partial t}\left[a\left(\dot K+WS\right)\right]-
K''~~~~~~~~~~~~~~~~~~~~~~~~~~~~~~~~\nonumber\\
+\left[\frac{\left(K^2-1+S^2 \right)K}{\xi^2}
+\left(\dot S-WK \right)W\right]=0.\nonumber
\end{eqnarray}

One can readily verify that the substitution  (\ref{9}) into equations (\ref{13}) yields the only equation for unknown function $P(r)$, namely
\begin{equation}\label{14}
P''-\frac{(P^2 -1)P}{\xi^2}= 0.
\end{equation}
At the same time, inserting (\ref{9}) into  (\ref{11}), one can obtain the following components of EMT for the YM field
\begin{eqnarray}
T_0^0 = \frac{1}{8\pi g^2 a^4 \xi^2}\left( 2 P'^2+\frac{(P^2
-1)^2}{\xi^2}\right),~~~~~~~\nonumber
\\
T_1^1 =\frac{1}{8\pi g^2 a^4 \xi^2}\left(- 2 P'^2+\frac{(P^2
-1)^2}{\xi^2}\right),~~~~\label{15}
\\
T_2^2 = T_3^3 = - \frac{1}{8\pi g^2 a^4}\frac{(P^2
-1)^2}{\xi^4},~~~~~~~~~~~~~~~\nonumber
\end{eqnarray}
and $T_0^1=0$.
Assuming  the displacement field $\phi_i(t)$ is a homogeneous  vector field, and considering the homogeneity and isotropy of the line element (\ref{2}), it is necessary to require the independence of the EMT components (\ref{15}) on the radial coordinate $r$. Thus, we get
\begin{eqnarray}
\left\{\begin{array}{rcl}
\displaystyle 2 P'^2+\frac{(P^2
-1)^2}{\xi^2}=A \xi^2,~~~~~~~~~~~~~~~~~~~~~~\\
\\
\displaystyle 2 P'^2-\frac{(P^2
-1)^2}{\xi^2}=B \xi^2,~~~~~~~~~~~~~~~~~~~~~\\
\\
\displaystyle \frac{(P^2
-1)^2}{\xi^2}=C\xi^2,~~~~~~~~~~~~~~~~~~~~~~~~~~~~~~~\\
\end{array}
\right.\label{16}
\end{eqnarray}
where $A$, $B$ and $C$ are some constants. A simple analysis of equations (\ref{14}) and (\ref{16}) allows to find out that $A=3C$ and $B=C$. Therefore, the set of equations (\ref{16}) is reduced to the following two equations
\begin{equation}
P'^2=C \xi^2,~~~~~
P^2=1+\epsilon \sqrt{C}\xi^2,\label{17}
\end{equation}
where $\epsilon = \pm 1$. The direct consequence of (\ref{17}) is as follows
\begin{equation}
\xi'^2=1+\epsilon \sqrt{C}\xi^2.\label{18}
\end{equation}
Taking into account (\ref{3}) and (\ref{17}), we can get $C=|k|$, $\epsilon=-k$ and
\begin{equation}
P(r)=\pm \sqrt{1- k \xi^2}.\nonumber
\end{equation}
Hence, due to (\ref{3}), the latter leads to
the nontrivial solution for equation (\ref{14}) given by
\begin{eqnarray}
P(r)=\pm \xi'(r)=\left\{\begin{array}{rcl}
\pm\cos r,~~~~&&k =+1,~~~~~~~~~\\
\pm 1,~~~~~~~~&&k =0,~~~~~~~~~~~~~\\
\pm \cosh r,~~&&k =-1.~~~~~~~~~\\
\end{array}
\right.
 \label {19}
\end{eqnarray}

Interpreting EMT of the YM gauge field as the EMT of a perfect fluid, that is
\begin{equation}\label{20}
T_{ik}= (\rho_g +p_g)u_i u_k -p_g\, g_{ik},
\end{equation}
we can get from (\ref{15}) and (\ref{19}) that
\begin{equation}\label{21}
T_0^0=\rho_g=\frac{3|k|}{8\pi g^2 a^4},~T_1^1=T_2^2=T_3^3=-p_g=- \frac{|k|}{8\pi g^2 a^4},
\end{equation}
Thus, the effective EoS of the given YM field is the same one as  for the pure radiation, that is $p_g=\rho_g/3$, when $k=\pm1$.

It is useful to mention that the conservation equation for YM field, $T^k_{i;k}=0$, for EMT of YM field in form (\ref{21}), namely
\begin{equation}\label{22}
\dot \rho_g+3H(\rho_g+p_g)=0,
\end{equation}
where $H=\dot a/a$ is the Hubble parameter, is satisfied identically by our solution (\ref{21}).
As can be seen, the reason of such satisfaction of the continuity equation in the case $k=0$ is nothing more than the equality of $\rho_g$ and $p_g$ to zero according to (\ref{21}).
That is why we will consider further only two types of the non-flat models with $k = \pm 1$.

%==============================
\section{Cosmological equations}
%==============================

Let $\phi_i$ be a time-like displacement vector field,
\begin{equation}\label{23}
\phi_i = \left(\frac{2}{\sqrt{3}}\,\beta,0,0,0\right),
\end{equation}
where $\beta = \beta(t)$ is a function of time only, and the factor $2/\sqrt{3}$ is used in order to simplify the writing of all the following equations.
Given the line element (\ref{2}) and  taking the total EMT as a sum of (\ref{7}) and (\ref{20}), one can reduce the main equations of the model (\ref{1}) to the following set of equations:
\begin{eqnarray}
3H^2 + \frac{3 k}{a^2} - \beta^2 = \rho_m + \rho_g +\Lambda,~~~~~~~~~~~~\label{24}
\\
2 \dot H + 3H^2 + \frac{k}{a^2} + \beta^2 = -  p_m -p_g +\Lambda.~~~\label{25}
\end{eqnarray}

The continuity equation follows from Eqs. (\ref{24}) and (\ref{25}) as
\begin{equation}\label{26}
\dot \rho_m + \dot \rho_g + \dot \Lambda + 2 \beta \dot \beta + 3 H \Big(\rho_m+\rho_g + p_m +p_g+ 2\beta^2 \Big)=0.
\end{equation}

Obviously, the displacement vector field can be treated through the EMT in the same way as in the most studies on the Lyra geometry in cosmology. In such approach, the conservation of the energy-momentum should be accounted for the sum of EMTs \cite{Hova}.  As emphasized in \cite{Shchigolev1}, the displacement vector field can give rise to an effective cosmological term $\Lambda_{eff}$. It means that $\Lambda(t)$ is not an independent dynamical parameter of the model, and it should be removed from the system of equations (\ref{24}) and (\ref{25}).

Moreover,  we are going to preserve the continuity equation for matter in its standard form,
\begin{equation}
\dot \rho_m + 3 H (\rho_m + p_m )=0,\label{27}
\end{equation}
which follows from the conservation equation, $\mathcal{T}^k_{i;k}=0$, and (\ref{7}).
Taking into account equations (\ref{20}) and (\ref{27}), the continuity equation  (\ref{26}) becomes
\begin{equation}\label{28}
\dot \Lambda + 2 \beta \dot \beta + 6 H \beta^2 =0.
\end{equation}
Since $\Lambda =constant$, equation (\ref{28}) supposes the displacement field to be the so called stiff fluid. The assumption of a non-vanishing and time-varying $\Lambda$ term gives us some new possibilities \cite{Shchigolev2}. Therefore, we assume that  $\Lambda \neq constant$. Moreover, we suppose that  (\ref{28}) can be satisfied by some $\Lambda(t)$, which can be found as the result of formal integration of equation (\ref{28}),
\begin{equation}
\Lambda(t) =\lambda_0 -\beta^2-6\int H \beta^2 d t, \nonumber
\end{equation}
where $\lambda_0$ is a constant of integration. Substituting this expression along with $\rho_g, p_g$ from (\ref{21}) into equations (\ref{24}) and (\ref{25}), one can get the basic equations of the model as follows
\begin{eqnarray}
3H^2 + \frac{3k}{a^2} = \frac{3}{8\pi g^2 a^4}+\rho_{eff},~~~~~~~~~~~~~\label{29}\\
2 \dot H -\frac{2k}{a^2} = -\frac{1}{2\pi g^2 a^4} - (\rho_{eff} + p_{eff}),~~~~~~~~\label{30}
\end{eqnarray}
where we have introduced the effective cosmological term
\begin{equation}\label{31}
\Lambda_{eff}(t)=\lambda_0 - 6 \int H \beta^2 d t,
\end{equation}
and the effective energy density and pressure
\begin{equation}\label{32}
\rho_{eff}=\rho_m +\Lambda_{eff},~~~~
p_{eff}=p_m -\Lambda_{eff} - \frac{\dot\Lambda_{eff}}{3 H}.
\end{equation}

One can readily verify that due to  equations (\ref{29}) and (\ref{30}), and the continuity equation (\ref{27}), the effective energy density and pressure (\ref{32}) also satisfy the continuity equation in its usual form:
\begin{equation}
\dot \rho_{eff} + 3 H (\rho_{eff} + p_{eff} )=0.\label {33}
\end{equation}

At the same time, the deceleration parameter in Lyra's geometry is defined just as in the standard cosmology, that is
\begin{equation}\label{34}
q = -\frac{a^2\, \ddot a}{\dot a^2} = -1-\frac{\dot H}{H^2}.
\end{equation}
It should be noted that the set of dynamical equations (\ref{29}), (\ref{30}) and (\ref{33}) consists of two independent equations, and fully determines the dynamics of our model. However, to determine three parameters, say $a(t)$, $\rho_{eff}$ and $p_{eff}$,  one more condition should be set. For example,  an effective EoS $w_{eff}=p_{eff}/\rho_{eff}$ can play the role of such  additional equation.
Assuming the matter also obeys  a barotropic EoS  $p_m = w_m \rho_m$, we also can obtain the full set  of equations but only if we make some assumption about $\Lambda_{eff}(t)$ (see, e.g. \cite{Matyjasek3}). Let us now consider one interesting example of exact solution for our model.

%
%=============================================
\section{Solution for the effective vacuum model}
%=============================================

In the following, we provide the exact solution
for our model in one simple case. Let us suppose
\begin{equation}\label{35}
p_{eff}=-\rho_{eff},
\end{equation}
that is the effective vacuum EoS $w_{eff}=-1$. From equation (\ref{33}), we get
\begin{equation}
\rho_{eff} = \Lambda_0=constant>0.\label{36}
\end{equation}
Hence, the only independent equation, which has to be solved, follows from (\ref{29}) as
\begin{equation}\label{37}
3H^2 + \frac{3k}{a^2} = \frac{3}{8\pi g^2 a^4}+\Lambda_0.
\end{equation}
Equation (\ref{37}) can be readily solved for all possible relations between its constant parameters (see, e.g., \cite{Stephani} and references therein). In particular, for a special choice of the constants of integration, these solutions can be given by $a=a_{k,\delta}(t)$,
\begin{eqnarray}
a\!=\!\left\{\begin{array}{rcl}
\!\!\!\!\!\!\left(\!\!\sqrt{A}\sinh{\Big[2\sqrt{\frac{\Lambda_0}{3}}(t+t_{ks})\Big]}+\frac{3k}{2\Lambda_0}\right)^{1/2}\!\!\!\!\!\!\!,~\delta=+1,\\ \\
\!\!\!\!\!a_0\,\exp\Big[\sqrt{\frac{\Lambda_0}{3}}t\Big],~~~~~~~~~~~~~~~~~~~~~~~~~~~\delta=0,\\ \\
\!\!\!\!\!\!\left(\!\!\sqrt{A}\cosh{\Big[2\sqrt{\frac{\Lambda_0}{3}}(t+t_{kc})\Big]}+\frac{3k}{2\Lambda_0}\right)^{1/2}\!\!\!\!\!\!\!,~\delta=-1,\\
\end{array}
\right.
\label {38}
\end{eqnarray}
where
\begin{equation}\label{39}
A=\frac{3}{4\Lambda_0}\left|\frac{1}{2\pi g^2}-\frac{3}{\Lambda_0}\right|,~~\delta=sgn\left(\frac{1}{2\pi g^2}-\frac{3}{\Lambda_0}\right),
\end{equation}
and the constants of integration  are supposed to be
\begin{equation}\nonumber
t_{ks}=\frac{(1-k)}{4}\sqrt{\frac{3}{\Lambda_0}}\text{arsinh}\left(\frac{3}{2\sqrt{A}\Lambda_0}\right)
\end{equation}
and
\begin{equation}\nonumber
t_{kc}=\frac{(1-k)}{4}\sqrt{\frac{3}{\Lambda_0}}\text{arcosh}\left(\frac{3}{2\sqrt{A}\Lambda_0}\right).
\end{equation}

Therefore, we can obtain from (\ref{38}) the following expression for the Hubble parameter $H(t)=H_{k,\delta}(t)$,
\begin{eqnarray}
H_{k,\delta}(t)=\left\{\begin{array}{rcl}
\displaystyle \!\!\!H_0\frac{\cosh{\big[2H_0(t+t_{ks})\big]}}{\sinh{\big[2H_0(t+t_{ks})\big]}+B_k},~~\delta=+1,\\ \\
\displaystyle \!\!\!\!H_0\frac{\sinh{\big[2H_0(t+t_{ks})\big]}}{\cosh{\big[2H_0(t+t_{ks})\big]}+B_k},~~\delta=-1,\\
\end{array}~
\right.
 \label {40}
\end{eqnarray}
where $H_0=\sqrt{\Lambda_0/3}$, $B_k=\displaystyle \frac{k}{2\sqrt{A}H_0^2}$, and $H(t)=H_0$ for the case of $\delta=0$. As one can see, at any case, $H_0 = \lim_{t \to \infty}H(t)$.

\begin{figure}
\centering
\includegraphics[width=0.85\columnwidth]{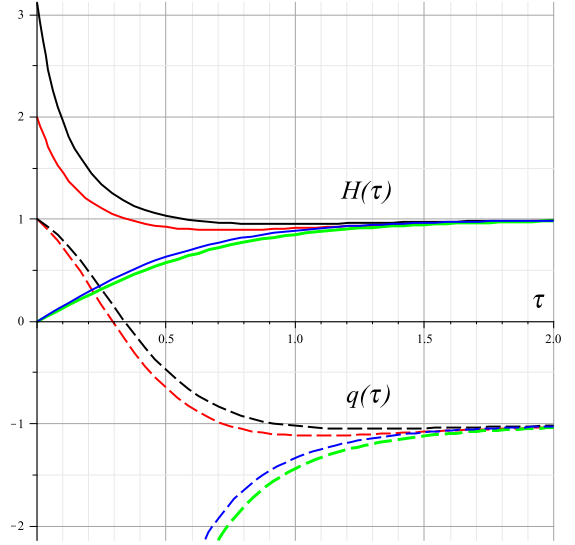}
\caption{Shows the Hubble parameter (\ref{40}) and the deceleration parameter (\ref{41}) versus the dimensionless time $\tau=H_0 t$.}
\end{figure}

Taking into account these equations and (\ref{34}), one can get the deceleration parameter $q=q(t)$ as follows
\begin{eqnarray}
q\!=\!\left\{\begin{array}{rcl}
\!\!\!\!\displaystyle \frac{B_k^2\!+\!1\!-\!\Big(\sinh{\big[2H_0(t+t_{ks})\big]}+ B_k\Big)^2}{\cosh^2\big[2H_0(t\!+\!t_{ks})\big]},~\delta\!=\!+1,\\ \\
\!\!\!\!\displaystyle  \frac{B_k^2\!-\!1\!-\!\Big(\cosh{\big[2H_0(t+t_{ks})\big]}+ B_k\Big)^2}{\sinh^2\big[2H_0(t\!+\!t_{ks})\big]},~\delta\!=\!-1.\\
\end{array}~
\right.
 \label {41}
\end{eqnarray}

The behavior of the  Hubble parameter (\ref{40}) and the deceleration parameter (\ref{41}) with time for $k=+1$ and different values of $\delta$ is shown in Fig. 1. To be specific, we have chosen $A=1$. Then, one can see the black lines for $H_0=1$ and $\delta=+1$, the blue lines for $H_0=1$ and $\delta=-1$, the red lines for $H_0=1.25$ and $\delta=+1$, and the green lines for $H_0=1.25$ and $\delta=-1$.

Applying (\ref{35}) for (\ref{32}) and (\ref{27}), we get
\begin{equation}\label{42}
p_m+\rho_m=\frac{\dot\Lambda_{eff}}{3 H},
\end{equation}
and, taking into account (\ref{35}) and (\ref{36}),
\begin{equation}\label{43}
\rho_m=\Lambda_{0}-\Lambda_{eff},~~~~p_m=-\Lambda_{0}+\Lambda_{eff} +\frac{\dot\Lambda_{eff}}{3 H}.
\end{equation}
The last means that the matter EoS is
\begin{equation}\label{44}
w_m=\frac{p_m}{\rho_m}=-1+\frac{\dot\Lambda_{eff}}{3 H(\Lambda_{0}-\Lambda_{eff})}.
\end{equation}

Since
\begin{equation}\label{45}
\dot \Lambda_{eff}= - 6 H \beta^2
\end{equation}
in accordance to (\ref{31}), than equation (\ref{42}) leads to
\begin{equation}\label{46}
p_m+\rho_m=-2\beta^2.
\end{equation}
Let us suppose hat the matter in our model exists in the form of a scalar field $\phi(t)$ with a potential $V(\phi)$, that is
\begin{equation}\label{47}
\rho_m=\frac{1}{2}\epsilon\dot \phi^2 + V(\phi), ~~~ p_m=\frac{1}{2}\epsilon\dot \phi^2 - V(\phi),
\end{equation}
where $\epsilon=\pm1$ for the quintessential and phantom fields, respectively.
Due to the real character of the displacement vector and non-negativity of the matter density,  we obtain  $w_m \le -1$. It means that the matter should be of the phantom nature. For example, it can be a phantom scalar field $\phi(t)$, for which $\epsilon=-1$, that is
\begin{equation}\label{48}
\rho_m+p_m=-\dot \phi^2,~~~ \rho_m-p_m= 2 V(\phi)
\end{equation}
From these expressions and equations (\ref{43}) - (\ref{46}), we have
\begin{equation}\label{49}
\dot \phi^2=2\beta^2,
\end{equation}
and
\begin{equation}\label{50}
V =  \Lambda_0+\beta^2-\Lambda_{eff}.
\end{equation}
The further investigation of this model involves the imposition of some additional conditions.
For example, we can supplement the model by the following simplest assumption regarding the displacement field
\begin{equation}\label{51}
\beta=\beta_0=constant.
\end{equation}
which is often discussed in the literature.
Taking into account equations (\ref{49}) and (\ref{51}), we obtain
\begin{equation}\label{52}
\phi(t)=\sqrt{2}\beta_0\,t+\phi_0,
\end{equation}
where $\phi_0$ is a constant of integration. In view of (\ref{51}), the integration of equation (\ref{45}) yields the following effective cosmological term
\begin{equation}\label{53}
\Lambda_{eff}= \lambda_0- 6 \beta_0^2\ln\left[\frac{a(t)}{a_0}\right],
\end{equation}
where $\lambda_0$ is a constant of integration. Substituting (\ref{53}) into (\ref{50}) alon with (\ref{51}) and (\ref{52}), we get the following potential of the scalar field
\begin{equation}\label{54}
V(\phi)= \overline{V}_0 + 6 \beta_0^2\ln \Big|a\left(\frac{\phi-\phi_0}{\sqrt{2}\beta_0}\right)\Big|,
\end{equation}
where $\overline{V}_0 = \Lambda_0+\beta_0^2-\lambda_0- 6 \beta_0^2\ln a_0$.
Substituting equation (\ref{38}) for different $\delta$ and $k=\pm1$ in (\ref{54}), we can obtain the scalar potential $V(\phi)=V_{k,\delta}(\phi)$ as follows
\begin{equation}\label{55}
V_{k,+1}(\phi)= V_0 + 3 \beta_0^2\ln \left|\sinh{\left(\frac{1}{\beta_0}\sqrt{\frac{2\Lambda_0}{3}}\,\phi\right)}+\frac{3k}{2\sqrt{A}\Lambda_0}\right|,
\end{equation}
for $\delta=+1$ and $\phi_0=\sqrt{2}\beta_0 t_{ks}$,
\begin{equation}\label{56}
V_{k,-1}(\phi)= V_0 + 3 \beta_0^2\ln \left|\cosh{\left(\frac{1}{\beta_0}\sqrt{\frac{2\Lambda_0}{3}}\,\phi\right)}+\frac{3k}{2\sqrt{A}\Lambda_0}\right|,
\end{equation}
for $\delta=-1$ and $\phi_0=\sqrt{2}\beta_0 t_{kc}$, and
\begin{equation}\label{57}
V(\phi)= V_0 + \sqrt{6\Lambda_0}\,\beta_0\,\phi,
\end{equation}
when $\delta=+1$. Here $V_0$ is a constant.
\begin{figure}
\centering
\includegraphics[width=0.85\columnwidth]{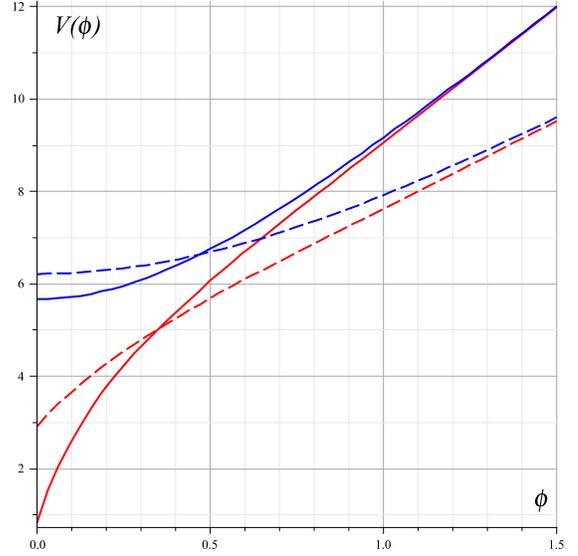}
\caption{Shows the scalar field potentials (\ref{55}) (in red) and (\ref{56}) (in blue) for $k=+1$, $\beta_0=A=1$, $V_0=5$,  and two different values of $\Lambda_0$: 6 (solid lines) and 3 (dashed lines).}
\end{figure}
For the specific values of constants, the scalar potentials (\ref{55}) and (\ref{56}) are shown in Fig.2.

Substituting (\ref{53}) in equation (\ref{44}), we get the following equation for the scalar field EoS
\begin{equation}\label{58}
w_m(t)=-1-\frac{1}{R+3\ln\left[\displaystyle\frac{a(t)}{a_0}\right]},
\end{equation}
where $R=(\Lambda_0-\lambda_0)(2\beta_0^2)^{-1}$, and $a(t)$ is defined by equation (\ref{38}).
As can be seen from equation (\ref{58}), the EoS of matter$w_m$ could remain less than $-1$ all the time but tends to $-1$ as $t\to\infty$.

%==============================
\section{Cosmological model with a constant Hubble parameter}
%==============================

Now we consider the case of inflationary model, when $H=H_0 = constant$, and hence
\begin{equation}\label{59}
a(t)=a_0 \exp(H_0 t).
\end{equation}
Then from equation (\ref{31}), one can obtain
\begin{equation}\label{60}
\Lambda_{eff}(t)=\lambda_0 - 6H_0\int \beta^2(t) d t .
\end{equation}

According to equations (\ref{29}), (\ref{32}) and (\ref{60}), it is easy to get the following energy density and pressure of matter

\begin{eqnarray}
\rho_m\!=\!\frac{3k}{a^2}\!-\!\frac{3}{8\pi g^2a^4}\!-\!\lambda_0+3H_0^2\!+6H_0\int \beta^2 d t,~~~~~~ \\
p_m \!=\!-\frac{k}{a^2}\!-\!\frac{1}{8\pi g^2a^4}\!+\!\lambda_0-3H_0^2\!-\!6H_0\int \beta^2 d t -\!2 \beta^2.\nonumber
\label{61}
\end{eqnarray}
By summarizing these equations, one can get that
\begin{equation}\label{62}
\rho_m + p_m = \frac{2k}{a^2}-\frac{1}{2\pi g^2a^4}-2\beta^2.
\end{equation}

Once again, the further investigation of this model requires to apply some extra conditions.
For example, we can supplement the model by the following simplest assumption regarding the displacement field
\begin{equation}\label{63}
\beta(t)= \frac{\beta_0}{a^2(t)}=\frac{\beta_0}{a_0^2}\exp(-2H_0\,t),
\end{equation}
where $\beta_0$ is a constant. Taking into account (\ref{59}), (\ref{61}) and  (\ref{63}), one can get the following equation for the EoS of matter $w_m=p_m/\rho_m$
\begin{equation}\label{64}
w_m =-\frac{\displaystyle\frac{k}{a_0^2}\,e^{-2H_0\,t}+\frac{1}{2a_0^4}\left(\frac{1}{4\pi g^2}+\beta_0^2\right)\,e^{-4H_0\,t}+\rho_0}{\displaystyle\frac{3k}{a_0^2}\,e^{-2H_0\,t}-\frac{3}{2a_0^4}\left(\frac{1}{4\pi g^2}+\beta_0^2\right)\,e^{-4H_0\,t}+\rho_0},
\end{equation}
where and further $\rho_0=3H_0^2-\lambda_0$ is the asymptotic (at $t\to \infty$) value of the energy density of matter. It is interesting to note that this equation asymptotically leads to the constant quintessential EoS  $w_m=-1/3$ when and only when $\rho_0=0$, that is $\lambda_0=3H_0^2$, for any sign of the curvature, and for any numeric values of other parameters of the model.
\begin{figure}
\centering
\includegraphics[width=0.85\columnwidth]{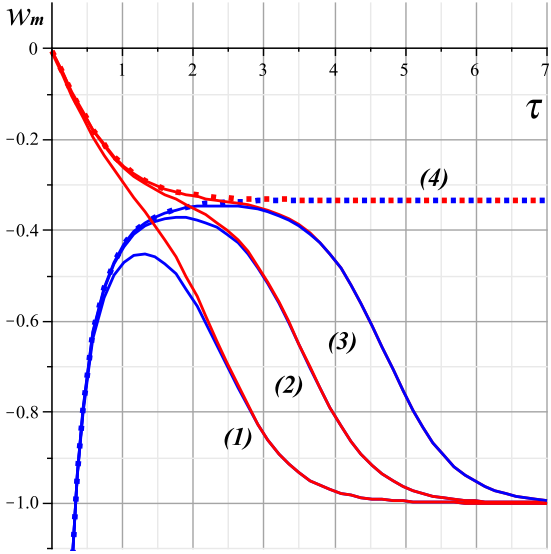}
\caption{Shows the dependence of $w_m(\tau)$ in (\ref{64}) on $\tau=H_0 t$ for $\rho_0$ equals (1) $0.01k$, (2) $0.001k$, (3) $0.0001k$ and  (4) $\rho_0=0$.}
\end{figure}
In general, the detailed behavior of $w_m(t)$ depends on the numeric values of parameters $k$, $a_0$, $\beta_0$ and $g$ in equation (\ref{64}), and also depends on the relation between them. If we introduce the following notation
\begin{equation}\label{65}
 C_1=\frac{2}{a_0^2},~~ C_2=\frac{2}{a_0^4}\left(\frac{1}{4\pi g^2}+\beta_0^2\right),
\end{equation}
we can find that the EoS parameter (\ref{64}) behaves itself  in  different manners depending to the relative relation between $C_1$ and $C_2$. The dependence of EoS $w_m(\tau)$  on the dimensionless time $\tau=H_0 t$ for  $C_1=0.8$, $S_2=1.6$ and $k=+1$ (in blue), $k=-1$ (in red) is shown in Fig. 3 for several values of $\rho_0$.

Let us assume again that the matter exists in the form of a scalar field $\phi(t)$ with a potential $V(\phi)$. Therefore, it follows from equation (\ref{47}) that
$$
\rho_m+p_m=\epsilon\dot \phi^2,~~~ \rho_m-p_m= 2 V(\phi),
$$
From these equations and (\ref{59})-(\ref{63}),  we can obtain
\begin{equation}\label{66}
\epsilon H_0^2 (\phi')^2 =kC_1-C_2x^2,
\end{equation}
and
\begin{equation}\label{67}
V(x) =  kC_1x^2-\frac{1}{4}C_2x^4 + \rho_0,
\end{equation}
where $x=\exp(-H_0\,t)\in (0,1]$ and $\phi'=d\phi/dx$.
The exact solutions for equation (\ref{66}) can be readily obtained as follows
\begin{eqnarray}\label{68}
\phi(x) =\pm \frac{1}{2H_0}\Big[x\sqrt{C_1-C_2x^2}~~~~~~~~~~~~~~~~~~~~ \nonumber\\
+\frac{C_1}{\sqrt{C_2}}\arcsin\Big(\sqrt{\frac{C_2}{C_1}}x\Big)\Big]+\phi_0,
\end{eqnarray}
for $k=+1$, $C_1\geq C_2$, $\epsilon=+1$, or  $C_1<C_2$, $x\in [0,x_0]$ where $x_0=\sqrt{C_1/C_2}$,
\begin{eqnarray}\label{69}
\phi(x) =\pm \frac{1}{2H_0}\Big[x\sqrt{C_2x^2-C_1}~~~~~~~~~~~~~~~~~~~~ \nonumber\\
-\frac{C_1}{\sqrt{C_2}} \text{arcosh}\Big(\sqrt{\frac{C_2}{C_1}}x\Big)\Big]+\phi_0,
\end{eqnarray}
for $k=+1$, $C_1<C_2$, $\epsilon=-1$, $x\in [x_0,1]$, and
\begin{eqnarray}\label{70}
\phi(x) =\pm \frac{1}{2H_0}\Big[x\sqrt{C_1+C_2x^2}~~~~~~~~~~~~~~~~~~~~ \nonumber\\
+\frac{C_1}{\sqrt{C_2}} \text{arsinh}\Big(\sqrt{\frac{C_2}{C_1}}x\Big)\Big]+\phi_0,
\end{eqnarray}
for $k=-1$, $\epsilon=-1$.

Notably that almost the same result could be obtained from the assumption of the other rate of changing for the displacement field, namely
\begin{equation}\label{71}
\beta(t)= \frac{\beta_0}{a(t)}=\frac{\beta_0}{a_0}\exp(-H_0\,t).
\end{equation}
In this case, equation (\ref{64}) for the EoS of matter is rewritten as follows
\begin{equation}\label{72}
w_m =-\frac{\displaystyle\frac{1}{a_0^2}(k-\beta_0^2)\,e^{-2H_0\,t}+\frac{1}{8\pi g^2a_0^4}\,e^{-4H_0\,t}+\rho_0}{\displaystyle\frac{3}{a_0^2}(k-\beta_0^2)\,e^{-2H_0\,t}-\frac{1}{8\pi g^2a_0^4}\,e^{-4H_0\,t}+\rho_0},
\end{equation}
and both equations (\ref{66}) and (\ref{67}) are valid again but with re-defined constants $C_1, C_2$ by
\begin{equation}\label{73}
 C_1=\frac{2(1-k\beta_0^2)}{a_0^2},~~ C_2=\frac{1}{2\pi g^2a_0^4}.
\end{equation}
Therefore, we can obtain the same solutions (\ref{68})-(\ref{70}) with new constants (\ref{73}).

Putting aside a detailed discussion of a generalization of this model in the spirit of Ref.\cite{Shchigolev3}, let us note that the consequences  of such generalization for the cosmic acceleration could be more radical than those obtained here.

\section{Conclusion}

In this paper, we have investigated some cosmological non-flat FRW models with YM fields on Lyra manifold. By using the exact solution of the YM equations on the FRW cosmological background, the exact solutions in Lyra's geometry have been obtained with  the certain propositions regarding the effective equation of state and the rate of expansion.

As a result, we have obtained the explicit expressions for the EoS of matter. The scalar field potential  has been reconstructed assuming that the matter  consists of the quintessential or phantom scalar field.
The results of this paper may also be extended without
much effort to the case of coupling the YM fields with the ordinary matter. In that case the accelerated regime of expansion could be reached by means of energy exchange between different forms of matter.

Finally, we can conclude that these models still contain a great potential for the study of cosmological expansion within Lyra geometry based on the exact solutions with participation of the Yang-Mills fields.

\end{document}